\documentclass[twocolumn,showpacs,prl,superscriptaddress]{revtex4}
\usepackage{amssymb}
\usepackage{amsmath}
\usepackage{graphicx}
\usepackage{epsfig}

\begin{document}

\title{Cooling a micro-mechanical resonator by quantum back-action from a
noisy qubit}
\author{Ying-Dan Wang}
\affiliation{NTT Basic Research Laboratories, NTT Corporation, 3-1 Morinosato Wakamiya,
Atsugi-shi, Kanagawa 243-0198, Japan}
\affiliation{Department of Physics, University of Basel, Klingelbergstrasse 82, 4056
Basel, Switzerland}
\author{Yong Li}
\affiliation{Department of Physics, University of Basel, Klingelbergstrasse 82, 4056
Basel, Switzerland}
\affiliation{Department of Physics, University of Hong Kong, Pokfulam Road, Hong Kong,
China}
\author{Fei Xue}
\affiliation{Department of Electrical Engineering, Technion, Haifa 32000, Israel}
\author{C. Bruder}
\affiliation{Department of Physics, University of Basel, Klingelbergstrasse 82, 4056
Basel, Switzerland}
\author{K. Semba}
\affiliation{NTT Basic Research Laboratories, NTT Corporation, 3-1 Morinosato Wakamiya,
Atsugi-shi, Kanagawa 243-0198, Japan}
\date{\today }

\begin{abstract}
We study the role of qubit dephasing in cooling a mechanical
resonator by quantum back-action. With a superconducting flux qubit
as a specific example, we show that ground-state cooling of a
mechanical resonator can only be realized if the qubit dephasing
rate is sufficiently low.
\end{abstract}

\maketitle

A micro (nano) mechanical resonator (MR) can be cooled down through
quantum back-action of coupled auxiliary mesoscopic
systems. Environmental fluctuations induce both relaxation and
dephasing processes to the auxiliary systems.  Previous
studies~\cite{Marquardt2007,Wilson2007,Genes2008,Wilson2004,
  Martin2004,Zhang2005,Hauss2008,Jaehne2008,Li2008,You2008,Grajcar2008}
show that relaxation plays an essential role to dissipate the MR
energy to the environment. In this paper, we address the function of
environment-induced dephasing in the back-action cooling.

Preparing quantum systems in their ground states is one way to
initialize them for the implementation of many quantum protocols. If
the MR is to be cooled to milli-Kelvin temperatures, novel cooling
techniques other than dilution refrigeration are required.
Experimentally, cooling by an optical cavity
(e.g.~\cite{Karrai2006,Kleckner2006,Poggio2007} and references
therein) and a superconducting single-electron transistor (SSET) has
been demonstrated~\cite{Naik2006}, but both techniques are still far
from reaching the quantum ground state of the MR. Theoretical
proposals have predicted the possibility of ground-state cooling of
the
MR~\cite{Marquardt2007,Wilson2007,Genes2008,Hauss2008,Jaehne2008,Li2008}.
The basic idea of these proposals resembles laser cooling of trapped
ions. The dissipative auxiliary system (such as the internal levels
of an ion, the optical cavity, the SSET, the quantum dot, etc.) acts
as a structured bath of the system to be cooled (in our case, the
MR). The relaxation and excitation rate of the auxiliary system
obeys the detailed balance relation determined by the external bath
temperature. Introducing a red-detuned drive modifies the
dissipative nature of the auxiliary system: absorption processes in
the auxiliary system and associated MR phonon emission processes
dominate over the inverse processes, so that detailed balance is
broken. These phonon emission processes extract energy from the MR
and dissipate it to the external bath through the subsequent
spontaneous emission of the auxiliary system energy quanta.
Therefore, the limit of the cooling procedure is determined
primarily by: (1) the environmental noise of the auxiliary system
and (2) the way the dissipative nature of the auxiliary system is
modified by the drive.

The environmental noise acting on the auxiliary system leads to both
relaxation processes (i.e., energy-lowering or energy-rising
processes) and dephasing processes. The cooling efficiency of
trapped-ion or optomechanical cooling is only influenced by the
relaxation process. In the case of cooling by driving a solid-state
qubit~\cite{Martin2004,Hauss2008,Wang2008,Jaehne2008}, pure
dephasing can usually be neglected by biasing the qubit at (or close
to) the degeneracy point where the pure dephasing rate is negligibly
small. However, the first-order photon excitation by absorbing the
energy from the low-frequency phonon and the linear drive vanishes
at this point, only a small second-order photon interaction
remains~\cite{Hauss2008}. Thus, the cooling becomes less efficient.
This motivates us to bias the qubit away from the degeneracy point
and take the qubit dephasing into account. This consideration is
especially important for solid-state qubits because the dephasing
rate is much larger than the relaxation rate away from the
degeneracy point and it is more difficult to suppress dephasing
since it is associated with low-frequency noise.

In this paper, we use the master-equation approach to investigate the
ground-state cooling of the MR by a flux qubit. The influence of both qubit
dephasing and relaxation is studied. Assuming a $1/f$ noise spectrum for the
flux qubit, we show that ground-state cooling of the mechanical resonator is
possible under current experimental conditions.

\begin{figure}[tp]
\begin{center}
\includegraphics[bb=106 450 490 570,scale=0.6,clip]{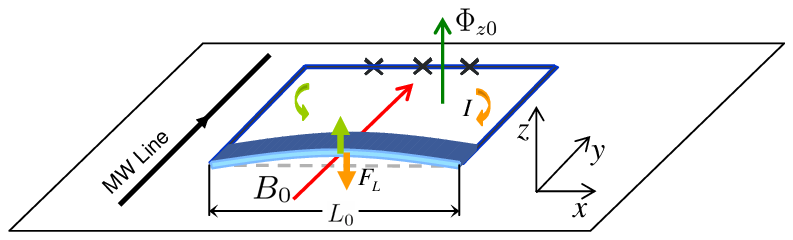}
\end{center}
\caption{(Color online) Schematic diagram of our setup. A doubly-clamped
mechanical beam is incorporated in a superconducting flux qubit (Josephson
junctions are indicated by crosses). The initial bias in the loop is
controlled by a magnetic flux $\Phi_{z0}$ in the $z$-direction. A coupling
magnetic field $B_{0}$ is applied in the $y$-direction, leading to a
coupling of the motion of the beam and the qubit because the supercurrent
leads to a Lorentz force on the beam. A microwave (MW) line introduces a
microwave bias to the qubit loop.}
\label{fig:circuit}
\end{figure}

The system we consider is a micro-mechanical beam interacting with a
superconducting flux qubit~\cite{Xue2007,Buks2006}, see
Fig.~\ref{fig:circuit}.
In the $x$-$y$ plane, a doubly-clamped micro-mechanical beam with an
effective length $L_{0}$ is incorporated in a superconducting loop
with three small-capacitance Josephson junctions. This mechanical
beam can be created from an MBE-grown heterostructure coated with
superconducting material~\cite{Etaki2008}, or a self-supporting
metallic air bridge. The fundamental vibration mode of the beam can
be well approximated as a harmonic resonator with oscillation
frequency $\omega_{\mathrm{b}}$. With a proper bias magnetic flux,
two classically stable states of the 3-Josephson-junction loop carry
persistent currents in opposite directions. There is a finite
tunneling rate $\Delta $ between the two classical
persistent-current states (throughout this article, we let
$\hbar=1$). This two-level subspace is far below the other energy
levels and forms a flux qubit~\cite{Mooij1999,Orlando1999}. The
qubit ground state $|g\rangle $ and excited state $|e\rangle $ are
coherent superpositions of two persistent current states denoted by
$|0\rangle $ (clockwise current state) and $|1\rangle $
(counter-clockwise current state). The energy spacing between the
two eigenstates is $\Omega =\sqrt{\Delta ^{2}+\varepsilon_{0}^{2}}$
where
$\varepsilon_{0}=2I_{\text{p}}\left( \Phi_{\text{ext}}-\Phi_{0}/2\right)$
is the energy spacing of the two classical current
states ($|0\rangle $ and $|1\rangle $), with $\Phi_{\text{ext}}$ the
external magnetic flux through the loop, $I_{\text{p}}$ the maximum
persistent current in the loop, and $\Phi_{0}=h/(2e)$ the flux
quantum. A microwave line is placed close to the circuit and can be
used as a microwave drive with frequency $\omega_{\mathrm{d}}$
acting on the flux qubit. The qubit Hamiltonian is written as
\begin{equation}
H_{\mathrm{q}}=(\Delta /2)\sigma_{x}
+(\varepsilon_{0}/2)\sigma_{z}+A\sigma_{z}\cos (\omega_{\mathrm{d}}t)
\end{equation}
where $A$ characterizes the amplitude of the microwave drive and
$\sigma_{x}\equiv \left\vert 1\right\rangle \left\langle
0\right\vert +\left\vert 0\right\rangle \left\langle 1\right\vert $,
$\sigma_{z}\equiv \left\vert 0\right\rangle \left\langle
0\right\vert -\left\vert 1\right\rangle \left\langle 1\right\vert$.
We assume that the drive is near resonant with the qubit $\left\vert
\delta \omega \right\vert /\left( \omega_{\mathrm{d}}+\Omega \right)
\ll 1$ (here, $\delta \omega =\Omega -\omega_{\mathrm{d}}$ is the
detuning between the qubit free energy and the drive), and that the
drive amplitude satisfies $A/\omega_{\mathrm{d}}\ll 1$. In the
presence of a magnetic field $B_{0}$ along the $y$-direction, the
supercurrent generates a Lorentz force $F_{\text{L}}$ on the MR
along the $z$-direction. This force couples the flux qubit with the
motion of the MR. The coupling Hamiltonian is
$g(a+a^{\dag }) \sigma_{x}$ with $g=B_{0}I_{p}L_{0}\delta_{0}$.
Here, $\delta_{0}=\sqrt{1/(2m\omega_{\mathrm{b}})}$ is the harmonic oscillator length
(mean-square zero-point displacement). As shown previously using a
semi-classical treatment~\cite{Wang2008}, this configuration can
serve as an ``on-chip refrigerator'' for the MR under a proper
drive: the Lorentz force produced by the flux qubit induces a
passive back-action on the MR, damping its thermal motion.

At low frequencies, the decoherence of the superconducting flux
qubit is dominated by $1/f$ noise. This noise induces qubit
relaxation and
dephasing~\cite{Yoshihara2006,Kakuyanagi2007,Bertet2005}. In the
absence of the microwave drive, the qubit relaxation (excitation)
rates $\Gamma_{\downarrow }^{(0)}$ , $\Gamma_{\uparrow }^{(0)}$
satisfy the detailed balance relation $\Gamma_{\downarrow
}^{(0)}/\Gamma_{\uparrow }^{(0)}=\exp (\Omega
/k_{\mathrm{B}}T_{0})$, where $T_{0}$ is the temperature of the
external bath. Since $\Omega \gg k_{\mathrm{B}}T_{0}$ in our case,
we neglect $\Gamma_{\uparrow }^{(0)}$ in this discussion. The qubit
pure dephasing rate $\Gamma_{\varphi }^{(0)}$ is almost
proportional to the qubit bias~\cite{Yoshihara2006,Kakuyanagi2007}
in the vicinity of the degeneracy point so that
\begin{equation}
\Gamma_{\varphi }^{(0)}=\alpha \varepsilon_{0}.
\end{equation}
In the presence of a near-resonant microwave drive, in the rotating frame of
frequency $\omega_{\mathrm{d}}$, there are new eigenstates which are
superpositions of the initial eigenstates $|e\rangle $ and $|g\rangle $. The
new eigenstates are the qubit-microwave dressed states~\cite{Cohenbook2}.
The dephasing of the undressed states also contributes to the relaxation and
excitation of the dressed states. The relaxation ($\Gamma_{\downarrow }$),
the excitation ($\Gamma_{\uparrow }$) and the dephasing rate
($\Gamma_{\varphi }$) of the dressed states can be generally written as
\begin{equation}
\Gamma_{\downarrow (\uparrow ,\varphi )}\equiv \Gamma_{\downarrow
(\uparrow ,\varphi )\text{R}}+\Gamma_{\downarrow (\uparrow ,\varphi
)\text{D}},  \label{rr}
\end{equation}
where $\Gamma_{...\text{R}}$ denotes the contribution from the relaxation
process and $\Gamma_{...\text{D}}$ from the dephasing process without
drive. The derivation of the Lindblad-form master equation of the driven
qubit in the rotating frame leads to the following explicit expressions for
the relaxation (excitation) rates:
\begin{eqnarray}
\Gamma_{\downarrow (\uparrow )\text{R}} &=&\frac{\Gamma_{\downarrow
}^{\left( 0\right) }\left( 1+\cos \phi \right) ^{2}}{4}\frac{\coth \left(
\beta \left( \omega_{0}+\omega_{\mathrm{d}}\right) /2\right) \pm 1}{\coth
\left( \beta \Omega /2\right) +1}  \notag \\
&&+\frac{\Gamma_{\downarrow }^{\left( 0\right) }\left( 1-\cos \phi
\right) ^{2}}{4}\frac{\coth \left( \beta \left( \omega_{0}-\omega
_{\mathrm{d}}\right) /2\right) \mp 1}{\coth \left( \beta \Omega
/2\right) +1}  \label{g1}
\\
\Gamma_{\downarrow (\uparrow )\text{D}} &=&\frac{\Gamma_{\varphi }^{\left(
0\right) }\sin ^{2}\phi }{2}\frac{\coth \left( \beta \omega_{0}/2\right)
\pm 1}{1+\coth \left( \beta \omega_{\mathrm{c}}/2\right) }, \\
\Gamma_{\varphi \text{R}} &=&\frac{\Gamma_{\downarrow }^{\left( 0\right)
}\sin ^{2}\phi }{\coth \left( \beta \Omega /2\right) +1} \\
\Gamma_{\varphi \text{D}} &=&\Gamma_{\varphi }^{\left( 0\right) }\cos
^{2}\phi  \label{g2}
\end{eqnarray}
Equations~(\ref{g1})-(\ref{g2}) are obtained by assuming a $1/f$
noise spectrum~\cite{Shnirman2002}. Here, $\beta
=1/(k_{\mathrm{B}}T_{0})$, $\sin \phi =A^{\prime }/\omega_{0}$,
$\cos \phi =\delta \omega /\omega_{0}$, $\omega_{0}=\sqrt{\delta
\omega ^{2}+A^{\prime 2}}$, and $A^{\prime }=-A\sin \theta $ with
$\sin \theta =\Delta /\Omega $ and
$\cos \theta =\varepsilon_{0}/\Omega $.
Below the cutoff frequency $\omega_{\mathrm{c}}$ of
the $1/f $ spectrum, we assume the noise spectrum to be white noise.

In the limit of weak qubit-resonator coupling, the Born approximation can be
applied and one can trace out the qubit degree of freedom in the composite
system master equation. This yields the following equation of motion for the
phonon number
\begin{equation}
\dot{n}=-n(\gamma_{\mathrm{m}}+\gamma_{\mathrm{q}})
+(\gamma_{\mathrm{m}}N+\gamma_{\mathrm{q}}N_{\mathrm{q}}) .
\end{equation}
Its solution reads
\begin{equation}
n\left( t\right) =n_{0}e^{-\left(
\gamma_{\mathrm{m}}+\gamma_{\mathrm{q}}\right)
t}+\frac{\gamma_{\mathrm{m}}N+\gamma_{\mathrm{q}}N_{\mathrm{q}}}{\gamma_{\mathrm{m}}+\gamma
_{\mathrm{q}}}\left( 1-e^{-\left(
\gamma_{\mathrm{m}}+\gamma_{\mathrm{q}}\right) t}\right)
\end{equation}
and the stable state solution at $t\gg
1/(\gamma_{\mathrm{m}}+\gamma_{\mathrm{q}})$ is
\begin{equation}
n=\frac{\gamma_{\mathrm{m}}N+\gamma_{\mathrm{q}}N_{\mathrm{q}}}{\gamma
_{\mathrm{m}}+\gamma_{\mathrm{q}}}.  \label{nf}
\end{equation}
Here, $\gamma_{\mathrm{m}}$ is the damping rate and $N=[\exp \left(
\beta \omega_{\mathrm{b}}\right) -1]^{-1}$ the mean phonon number determined
by the thermal bath, whereas
$\gamma_{\mathrm{q}}=g^{2}[S_{zz}(\omega_{\mathrm{b}}) -
S_{zz}(-\omega_{\mathrm{b}})] $ is the damping rate and
$N_{\mathrm{q}}=S_{zz}\left( -\omega_{\mathrm{b}}\right) /\left( S_{zz}\left(
\omega_{\mathrm{b}}\right) -S_{zz}\left( -\omega_{\mathrm{b}}\right)
\right) $ the mean phonon number determined by the qubit bath. Thus,
the MR is effectively in contact with two baths: One is the real
(external) thermal bath, and the other one is the structured bath
formed by the dissipative qubit under drive (for simplicity, we call
this the ``qubit bath'' in the following). Since the MR is coupled
via the $\sigma_{z}$-component of the qubit, the MR decoherence
related to the qubit bath is determined by the pair correlation
function $S_{zz}(\omega )=\int \mathrm{\mathrm{d}}te^{i\omega
t}\left\langle \sigma_{z}\left( t\right) \sigma_{z}\left( 0\right)
\right\rangle $, where the brackets denote the average over the
steady state of the qubit that satisfies $d\langle
{\sigma_{x,y,z}}\rangle /dt=0$. An explicit evaluation of
$S_{zz}(\omega )$ leads to
\begin{eqnarray}
S_{zz}\left( \omega \right) &=&\cos ^{2}\theta \sin ^{2}\phi \left( 1+\frac{
\kappa_{-}}{\kappa_{+}}\right) \frac{2\kappa_{0}}{2\left( \omega
+\omega_{0}\right) ^{2}+\kappa_{0}^{2}}+  \notag \\
&&\cos ^{2}\theta \sin ^{2}\phi \left( 1-\frac{\kappa_{-}}{\kappa
_{+}}\right) \frac{2\kappa_{0}}{4\left( \omega -\omega_{0}\right)
^{2}+\kappa_{0}^{2}}.  \label{sx}
\end{eqnarray}
Here, $\kappa_{0}=\Gamma_{\uparrow }+\Gamma_{\downarrow }+
2\Gamma_{\varphi}$,
$\kappa_{+}=\Gamma_{\uparrow}+\Gamma_{\downarrow}$, and
$\kappa_{-}=\Gamma_{\uparrow }-\Gamma_{\downarrow }$. Note that the
condition to obtain this reduced master equation is the weak
coupling between the MR and the two baths, i.e.
$\gamma_{\mathrm{m}}+\gamma_{\mathrm{q}}\ll \Gamma_{\downarrow
}^{(0)},\omega_{\mathrm{b}}$.

\begin{figure}[tp]
\begin{center}
\includegraphics[bb=112 274 513 573, scale=0.55,clip]{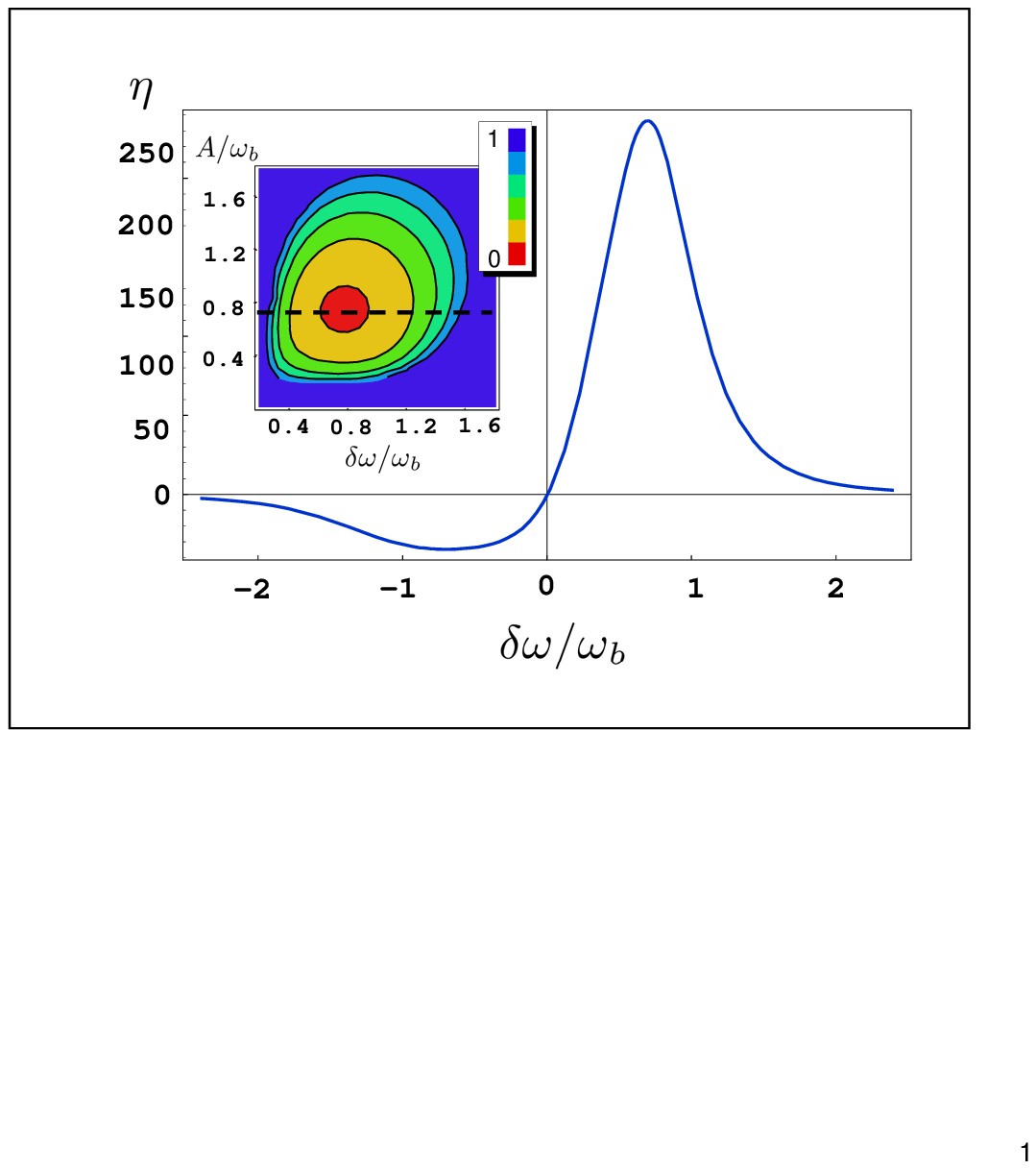}
\end{center}
\caption{(Color online) Inset: dependence of the final phonon number
$n$ on the scaled detuning
$\delta\omega/\omega_{\mathrm{b}}$ and the
scaled drive amplitude $A/\omega_{\mathrm{b}}$ (with
$\varepsilon_{0}=800$ MHz). The plot range is limited from
$n=0$ to $ n=1$ (i.e. the ground-state cooling regime). The region
with $n>1$ is shown in blue. Main panel: cooling power $\eta
\equiv N/n-1$ as a function of the scaled detuning
$\delta\omega/\omega_{\mathrm{b}}$ between
the microwave drive and qubit energy splitting along the dashed line
in the inset. The other parameters are the same as those used to
estimate the cooling limit in the text.} \label{fig:etadelta}
\end{figure}

In the absence of the drive, $\gamma_{\mathrm{q}}=0$, and
Eq.~(\ref{nf}) shows that the final occupancy number of the MR is
unchanged ($n=N$). This is a natural consequence of the second law
of thermodynamics. A slightly detuned drive breaks the thermal
equilibrium and changes the final phonon number of the MR.
Figure~\ref{fig:etadelta} shows the dependence of the cooling
efficiency $\eta \equiv N/n-1$ on the detuning $\delta \omega $. It
illustrates that cooling ($\eta >0$) is induced by a red-detuned
($\delta\omega>0$) drive: the red-detuned drive accelerates the
photon emission process so that the MR energy is dissipated into the
environment in the subsequent photon emission. In this way the mean
phonon number of the MR is decreased.

As an example, suppose we perform this cooling procedure on a MR of
length $L_{0}=5$ $\mu $m with oscillation frequency
$\omega_{\mathrm{b}}=10$ MHz, spring constant $k=0.01$ N/m, and
quality factor $Q_{\text{M}}=10^{4}$. The MR is coupled with a flux
qubit loop by an in-plane magnetic field $B_{0}=6$ mT. The qubit
energy splitting is $\Delta =5$ GHz at the degeneracy point and
$I_{\text{p}}=400$ nA. The coupling strength between the qubit and
the MR is about $10.2$ MHz. The initial environmental temperature is
assumed to be $T_{0}=20$ mK. The qubit relaxation rate
$\Gamma_{\downarrow }^{(0)}\approx 4$ MHz and the excitation rate
$\Gamma_{\uparrow }^{(0)}\approx 0$ are almost independent of the
qubit bias near the degeneracy point while the pure dephasing rate
depends linearly on the qubit bias in this regime so that
$\Gamma_{\varphi }^{(0)}\approx
0.008\varepsilon_{0}$~\cite{Yoshihara2006,Kakuyanagi2007}. The
cutoff frequency $\omega_{\mathrm{c}}$ is assumed to be $1$ Hz. If
the qubit is biased at $\varepsilon_{0}=1$~GHz and driven by a
microwave drive with $A=7.4$~MHz, $\omega_{\mathrm{d}}=5.056$~GHz,
the effective damping rate and the mean photon number of the qubit
bath are $\gamma_{\mathrm{q}}=0.4$ MHz, $N_{\mathrm{q}}=0.06$
respectively. The steady state is reached after about $2.5$ $\mu $s.
The phonon number of the MR in the steady state is $n=0.12$. Hence,
ground-state cooling can be realized with this setup at the
$1/f$-noise level characterized by the parameters
$\Gamma_{\downarrow }^{(0)}$, $\Gamma_{\uparrow }^{(0)}$,
$\Gamma_{\varphi }^{(0)}$, and $\omega_{c}$ given above which
correspond to realistic values. The cooling power can be further
improved by increasing the coupling magnetic field $B_{0}$.

In order to achieve ground-state cooling, the power and detuning of
the microwave drive should be optimized. As shown in the inset of
Fig.~\ref{fig:etadelta}, the final phonon number $n$ (or the cooling
power $\eta =N/n-1$) does not depend monotonically on the drive
magnitude. This is different from optomechanical cooling where a
stronger drive leads to a stronger cooling effect. Qubit-assisted
cooling requires a certain resonant condition and the qubit
eigen-frequency in the rotating frame is modified by the drive.
\begin{figure}[tp]
\begin{center}
\includegraphics[bb=89 1 377 185,scale=0.8,clip]{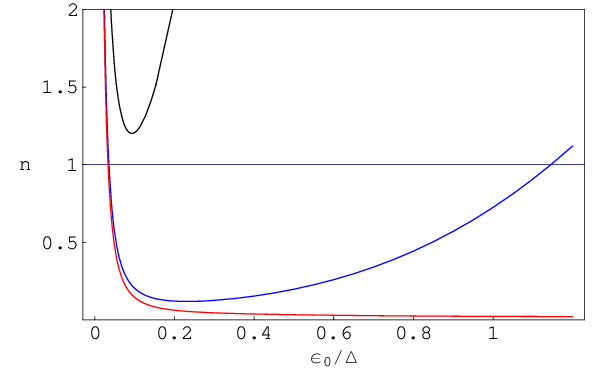}
\end{center}
\caption{(Color online) Final phonon number of the MR versus scaled
qubit bias. The three curves correspond to qubits with the same
relaxation rate ($\Gamma_{\downarrow }^{\left( 0\right) }\approx 4$
MHz) but different dephasing rates (from bottom to top
$\alpha =0$ (red), $\alpha =0.008$ (blue), and
$\alpha =0.08$ (black)).} \label{fig:neps0}
\end{figure}
Previous experiments on cooling a MR by quantum back-action~\cite{Naik2006}
show that the cooling limit is determined by the quantum fluctuations of the
auxiliary system. In the following, we analyze how the qubit dephasing and
relaxation influences the cooling process in a different way.

Since the decoherence of the qubit is sensitive to the qubit bias,
the dependence of the final phonon number on the qubit bias shown in
Fig.~\ref{fig:neps0} reveals the relationship between cooling
efficiency and qubit decoherence. Without consideration of dephasing
($\alpha=0$), the cooling efficiency increases monotonically with
the qubit bias. Taking a finite dephasing into account, there is an
optimal point for the cooling efficiency as we increase the qubit
bias. The cooling efficiency is decreased as the dephasing rate is
increased.

The behavior exhibited in Fig.~\ref{fig:neps0} can be understood
from Eq.~(\ref{sx}). Increasing dephasing broadens the Lorentzians
and decreases the difference between $\Gamma_\uparrow$ and
$\Gamma_\downarrow$. The correlation spectrum $S_{zz}(\omega)$ hence
becomes more symmetric and the cooling effect is decreased.
Therefore, $N_{\mathrm{q}}$ increases with the increase of pure
dephasing. Biasing the qubit to the degeneracy point helps to
decrease qubit dephasing and hence decrease $N_{\mathrm{q}}$.
However, at the degeneracy point, $\gamma_{\mathrm{q}}=0$, photon
excitation processes which absorb energy from the low-frequency
resonator and the linear drive are not possible any more and the
cooling cycle is stopped. Hence, there is an optimal qubit bias
point that leads to a lowest final phonon number, see
Fig.~\ref{fig:neps0}.

The dressed-state relaxation and excitation rates have contributions
from the qubit relaxation as well as pure dephasing. Physically,
this is because the dressed states of the qubit in the presence of
the microwave driving field are superpositions of the qubit
eigenstates in the absence of the drive. The qubit pure dephasing
contributes significantly to the decoherence process of the dressed
qubit and hence modifies the cooling limit. The pure dephasing was
not included in the previous studies of the trapped-ion cooling as
well as the cooling of a MR by a quantum dot, optical cavity, or
SSET. In these cases, the drive modifies the relaxation rate of the
auxiliary system but leaves the order of the eigenlevels unchanged.
The final relaxation process is not influenced by the pure
dephasing. In the cooling schemes that use a driven charge qubit,
dephasing processes are often neglected by biasing the qubit at the
degeneracy point. However, as shown in Fig.~\ref{fig:neps0}, at the
optimal bias point ($\varepsilon_{0}\approx 1$~GHz), the pure
dephasing rate $\Gamma_\varphi^{(0)}=8$~MHz is larger than the
relaxation rate $\Gamma_\downarrow^{(0)}=4$~MHz. Hence, it is
important to include dephasing in the study of ground-state cooling.
Ground-state cooling of a MR can only be realized if the qubit
dephasing rate is sufficiently low, see Fig.~\ref{fig:neps0}.

In conclusion, we have discussed the quantum theory of the cooling of a MR
using the back-action of a superconducting flux qubit. We have shown that
the present noise level of the qubit allows ground-state cooling of the MR.
The approach used in this paper can be applied to other qubit-assisted
back-action cooling schemes to estimate the cooling limit under the
influence of pure dephasing. Our method can also be applied to different
qubit noise spectra.

The authors are grateful to S. Saito, K. Kakuyanagi, and P. Zhang for
helpful discussions. This work has been partially supported by the JSPS
KAKENHI No. 18201018, the EC IST-FET project EuroSQIP, the Swiss SNF, and
the NCCR Nanoscience. F.X. was supported in part by the Aly Kaufman
Foundation.


\end{document}